\documentclass[11pt,a4paper]{article}
\usepackage{amsmath,amssymb}
\usepackage{epsfig,graphicx}
\usepackage{cite}

\begin{document}

\title{\bf Dirac neutrino magnetic moment and a possible time evolution 
of the neutrino signal from a supernova}

\author{R.~A.~Anikin\footnote{{\bf e-mail}: anik-roman@mail.ru},
A.~V.~Kuznetsov\footnote{{\bf e-mail}: avkuzn@uniyar.ac.ru},
N.~V.~Mikheev\footnote{{\bf e-mail}: mikheev@uniyar.ac.ru}
\\
\small{\em Yaroslavl State P.G.~Demidov University} \\
\small{\em Sovietskaya 14, 150000 Yaroslavl, Russian Federation}
}
\date{}

\maketitle

\begin{abstract}
We analyze the influence of neutrino helicity conversion, $\nu_L \to \nu_R$, on the neutrino flux from a
supernova, caused by the interaction of the Dirac neutrino magnetic moment with a magnetic field. We
show that if the neutrino has a magnetic moment in the interval 
$10^{-13} \, \mu_{\rm B} < \mu_\nu < 10^{-12} \, \mu_{\rm B}$ and provided that
a magnetic field of $\sim 10^{13} - 10^{14}$ G exists in the supernova envelope, a peculiar kind of time evolution of the neutrino signal from the supernova caused by the resonance transition $\nu_L \to \nu_R$ in the magnetic field
of the envelope can appear. If a magnetar with a poloidal magnetic field is
formed in a supernova explosion, then the neutrino signal could have a pulsating behavior, i.e., a kind of
a neutrino pulsar could be observed, when it rotates around an axis that does not coincide with its magnetic
moment and when the orientation of its rotation axis is favourable for our observation. 
\end{abstract}

\section{Introduction}
\label{sec:Introduction}


The processes of helicity flip, $\nu_L \longleftrightarrow \nu_R$, are possible
for a nonzero neutrino magnetic moment. In
the case of a Dirac neutrino, their realization under
conditions of a magnetized plasma in astrophysical
objects can be an important factor for the mechanism
of energy losses by such objects. In the standard
model extended to include the neutrino mass $m_\nu$, the
neutrino magnetic moment is known to be~\cite{Lee:1977,Fujikawa:1980} 
\begin{equation}
\mu_\nu^{(SM)} = \frac{3e\,G_{\rm F} \,m_\nu}{8\pi^2\sqrt{2}} 
= 3.20 \times 10^{-19} \left(\frac{m_\nu}{1 \,\textrm{eV}}\right)
\mu_{\rm B} \,,
\label{eq:mu_nu^SM}
\end{equation}
where $\mu_{\rm B} = e / 2 m_e$ is the Bohr magneton\footnote{We use a natural system of units with $c = \hbar = 1$. $e > 0$ is the elementary charge.}. Given
the existing constraints on the neutrino masses, this
quantity may be considered unobservably small. On
the other hand, various nontrivial extensions of the
standard model, such as the broken left-right symmetry~\cite{Lipmanov:1967,Lipmanov:1968a,Lipmanov:1968b,Pati:1974,Beg:1977}, admit
considerably larger neutrino magnetic moments~\cite{Kim:1976,Marciano:1977,Beg:1978}.

Considerable interest in the neutrino magnetic
moment arose after the momentous event of the
SN 1987A explosion~\cite{Hirata:1987,Hirata:1988,Bionta:1987,Bratton:1988,Alexeyev:1987,LSD:1987} in connection with the modeling
a supernova explosion in which the huge outgoing
neutrino flux essentially determines the energetics of
the process. This means that such a microscopic neutrino
characteristic as the magnetic moment could
have a decisive effect on the macroscopic properties
of these astrophysical events.

Two possible mechanisms for the realization of
neutrino helicity flip $\nu_L \longleftrightarrow \nu_R$ in astrophysical conditions
are discussed in the literature.

(1) The {\it scattering mechanism} is caused by 
the interaction of the Dirac neutrino magnetic moment
with the microscopic electromagnetic field of
a virtual plasmon. For example, interacting with the
plasmon that can be both produced and absorbed, the
trapped left-handed neutrinos in a supernova core can
be converted into right-handed ones:
\begin{eqnarray}
\nu_L \to \nu_R + \gamma^*, \quad \nu_L + \gamma^* \to \nu_R \, .
\label{eq:conversion}
\end{eqnarray}
These right-handed neutrinos are sterile with respect
to weak interactions, which can be important, for example,
when the energy losses by stars are taken into
account. The overly large flux of right-handed neutrinos
produced in such interactions from the supernova
core may not leave a sufficient amount of energy to
explain the observed supernova neutrino luminosity.
Thus, an upper bound on the neutrino magnetic
moment can be established (see, e.g.,~\cite{Barbieri:1988,Ayala:1999,Ayala:2000}; for a more
detailed list, see~\cite{Kuznetsov:2007}). The
contribution from the neutrino helicity flip process to
the supernova core luminosity was investigated most
consistently in our recent paper~\cite{Kuznetsov:2009a}. Here, instead of the model of a homogeneous
sphere used in previous studies, we considered realistic
models with radial distributions and time evolution
of physical parameters in the supernova core. We
obtained upper limits on the flavor-averaged Dirac
neutrino magnetic moment from the condition that
the influence of right-handed neutrino emission on
the total cooling timescale should be limited,
\begin{eqnarray}
\bar \mu_\nu < (1.1 - 2.7) \, \times 10^{-12} \, \mu_{\mathrm{B}}\,,
\label{eq:mu_lim_summ}
\end{eqnarray}
depending on the explosion model.

(2) The {\it mechanism of oscillations} $\nu_L \leftrightarrow \nu_R$ can
be realized when the neutrino magnetic moment interacts
with a macroscopic magnetic field in a supernova
envelope. The outgoing flux of right-handed
neutrinos from the core during collapse falls into the
region of the supernova envelope between the neutrinosphere
(of radius $R_\nu$) and the shock stagnation
zone (of radius $R_s$). According to existing views,
typical values of these quantities change insignificantly
in the stagnation time and can be estimated
as $R_\nu \sim 20$--$50$\,km and $R_s \sim 100$--$200$\,km. If a sufficiently
strong magnetic field, of the order of its critical
value of $B_e = m_e^2/e \simeq 4.41 \times 10^{13}$\,G, is present in
the region under consideration, then neutrino spin
oscillations take place.

The interesting possibility of a combination of both
these mechanisms was first suggested by Dar~\cite{Dar:1987},
who considered the process of double neutrino helicity
conversion, $\nu_L \to \nu_R \to \nu_L$, under supernova
conditions, where the first stage is realized through
the interaction of the neutrino magnetic moment with
plasma electrons and protons in the supernova core
and the second stage arises from the neutrino spin flip
in the magnetic field of the envelope. Voloshin~\cite{Voloshin:1988}
additionally took into account the possibility of
resonant conversion of right-handed neutrinos into
left-handed ones, $\nu_R \to \nu_L$, that was not considered
by Dar~\cite{Dar:1987}. In our recent paper~\cite{Kuznetsov:2009b}, we reanimated the idea of Dar~\cite{Dar:1987}
based on a new refined estimate~\cite{Kuznetsov:2007} for the flux and luminosity of right-handed
neutrinos from the central part of a supernova
(the previously used data were underestimated significantly).
We determined the conditions under which
the mechanism of double neutrino helicity conversion
could stimulate a damped shock during a supernova
explosion.

Here, we discuss a possibility of a combined action of a magnetic field and medium in the SN envelope on the outgoing neutrinos which could cause the resonant transition $\nu_L \to \nu_R$, and thus the SN neutrino signal could be modified. In principle, this effect could be observable.  
This talk is based on our recent paper~\cite{Anikin:2010}. 

\section{Neutrino helicity flip in a weakly magnetized plasma}
\label{sec:Helicity_flip}

It is most convenient to illustrate the influence of a
magnetic field on a neutrino with a magnetic moment
using the equation for neutrino helicity evolution in an
external magnetic field. Given the additional energy
acquired by the left-handed electron neutrinos $\nu_e$ in
the supernova envelope, the helicity evolution equation
can be written as~\cite{Voloshin:1986a,Voloshin:1986b,Okun:1986,Voloshin:1986c,Okun:1988,
Voloshin:1988,Blinnikov:1988}
\begin{equation}
{\mathrm i}\,\frac{\partial}{\partial t}
\left( 
\begin{array}{c} 
\nu_R \\ \nu_L 
\end{array} 
\right)
=
\left[\hat E_0 +
\left( 
\begin{array}{cc} 
0 & \mu_\nu B_{\perp} \\ \mu_\nu B_{\perp} & C_L
\end{array} 
\right)
\right]
\left( 
\begin{array}{c} 
\nu_R \\ \nu_L 
\end{array} 
\right) \,,
\label{eq:evolution} 
\end{equation}
where
\begin{equation}
C_L = \frac{3 \, G_{\mathrm F}}{\sqrt{2}} \, \frac{\rho}{m_N} 
\left( Y_e - \frac{1}{3} \right) \,.
\label{eq:C_L}
\end{equation}
Here, the ratio $\rho/m_N = n_B$ is the nucleon number
density, $Y_e = n_e/n_B = n_p/n_B$, $n_{e,p}$ are the electron
and proton number densities respectively, and $B_{\perp}$ is
the transverse magnetic field component with respect
to the direction of neutrino motion; the term $\hat E_0$ is
proportional to a unit matrix and is unimportant for
our analysis.

It should be explained why we use Eq.~(\ref{eq:C_L}) for the
additional energy of the left-handed electron neutrinos
in an unpolarized medium, although, in general,
at least a partial polarization of electrons should arise
in a field of the order of $B_e$. In this case, the validity of
the unpolarized-medium approximation is seen from
the following considerations. As is well known, the
states of electrons in a magnetic field corresponding
to all Landau levels, except the ground level, are
doubly degenerate with respect to the spin projection onto
the field direction and, thus, do not contribute to the
polarization of the medium. Therefore, to estimate the
degree of polarization, it will suffice to estimate the
fraction of electrons populating the ground Landau
level whose spins are uncompensated. For typical
conditions of the supernova envelope region under
consideration, the electron chemical potential is $\tilde \mu_e \simeq$
5-10 MeV (see, e.g.~\cite{Buras:2005}). Hence, dividing
the number density of the electrons populating
the ground Landau level, $n_0 \simeq e B \tilde \mu_e / (2 \pi^2)$, by
the total electron number density, $n_e \simeq \tilde \mu_e^3 / (3 \pi^2)$, we
obtain an estimate for the degree of polarization of the
medium:
\begin{equation}
P \sim \frac{n_0}{n_e} \lesssim \frac{e B}{\tilde \mu_e^2} 
\sim 10^{-2} \, \frac{B}{B_e} \,.
\label{eq:polaris}
\end{equation}
Thus, for the magnetic field strengths $B \sim B_e$ under
consideration, using the unpolarized-medium approximation
is justified. A more rigorous condition of
weak plasma magnetization under which the influence
of the magnetic field on the polarization of the
medium can be neglected is formulated as
\begin{equation}
B \ll \frac{(3 \pi^2 \, n_e)^{2/3}}{e} \simeq 0.6 \times 10^{16} \, \textrm{G} 
\left( \frac{n_e}{10^{33}\, \textrm{cm}^{-3}} \right)^{2/3} .
\label{eq:cond_B}
\end{equation}
Expression~(\ref{eq:C_L}) for the additional energy of the left-handed
neutrinos $C_L$ deserves a special analysis. A
remarkable possibility is that this quantity becomes
zero precisely in the supernova envelope region of
interest to us~\cite{Voloshin:1988}. This, in turn, is a
condition for the resonance transition $\nu_R \to \nu_L$ in the
form $Y_e = 1/3$. It should be noted that the values
of $Y_e$ typical of collapsing matter, $Y_e \sim$ 0.4-0.5, are
realized in the supernova envelope. However, causing
the dissociation of heavy nuclei, the shock wave
makes the matter more transparent to neutrinos. This
leads to the so-called ``short'' neutrino burst and, as
a consequence, to considerable deleptonization of the
matter in this region. According to existing views, a
characteristic dip where $Y_e$ can drop to $\sim 0.1$ arise
in the radial distribution of $Y_e$ (see, e.g.~\cite{Bethe:1990,Buras:2005}). Thus, there inevitably exists a
point where $Y_e$ takes on a value of 1/3. Remarkably,
there is only one such point with $\mathrm{d} Y_e / \mathrm{d} r > 0$, see~\cite{Bethe:1990,Buras:2005}.

Note that $Y_e = 1/3$ is a necessary but still insufficient
condition for resonant conversion of right-handed
neutrinos into left-handed ones, $\nu_R \to \nu_L$. The
adiabaticity condition should also be met. It means
that when shifted from the resonance point to a distance
of the order of the oscillation length, the diagonal
element $C_L$ in Eq.~(\ref{eq:evolution}) at least should not exceed
considerably the nondiagonal element $\mu_\nu B_{\perp}$.

\section{Time evolution of the neutrino flux}
\label{sec:Time_evolution}

The process of helicity flip for a Dirac neutrino with
a magnetic moment, which can lead to interesting observational
consequences when the expected neutrino
signal from an imminent supernova explosion is studied
in detail, can be realized. According to existing
views, during the explosion of a Galactic supernova
at a distance up to 10 kpc, the expected number of
neutrino events in the Super-Kamiokande detector
will be $\sim 10^4$. This will allow the time evolution of the
neutrino flux to be recorded with a good accuracy.

In the presence of a sufficiently strong magnetic
field in the supernova envelope, not only the above-mentioned
conversion of right-handed neutrinos
into left-handed ones, $\nu_R \to \nu_L$~\cite{Dar:1987,Voloshin:1988}, but also the conversion of active electron
neutrinos and antineutrinos of the main neutrino flux
into a form sterile with respect to weak interactions,
$\nu_L \to \nu_R, \, {\bar\nu}_R \to {\bar\nu}_L$, is possible.

Numerical analysis of Eq.~(\ref{eq:evolution}) shows that after its
passage through the resonance region ($Y_e = 1/3$),
the flux of left-handed neutrinos is attenuated as a
result of the above conversion by the factor $W_{LL}$,
which has the meaning of the survival probability of
left-handed neutrinos, $\nu_{eL} \to \nu_{eL}$, or, in other words,
the transparency. Figure 1 shows the characteristic
variation in $W_{LL}$ when passing through the resonance
point (placed here at the coordinate origin) for various
magnetic field strengths. We see that the supernova
envelope in the presence of a sufficiently strong magnetic
field is virtually opaque to active electron neutrinos
and antineutrinos, which can cause the expected
neutrino signal from the supernova to be attenuated.

\begin{figure}
\begin{center}
\includegraphics*[width=0.8\textwidth]{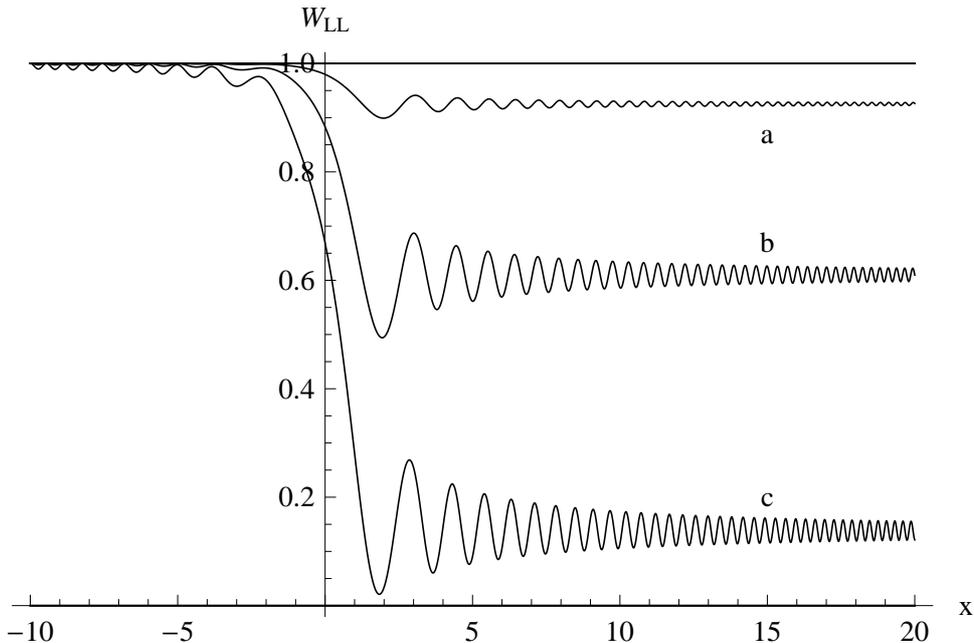}
\end{center}
\caption{Pattern of variations in $W_{LL}$, the survival probability of left-handed neutrinos, $\nu_{eL} \to \nu_{eL}$, (transparency), with distance $x$
(in arbitrary units) when passing through the resonance point placed at the coordinate origin for several magnetic field
strengths: $B = 0.2 \, B_e$ (a); $B = 0.5 \, B_e$ (b); $B = B_e$ (c). To be specific, the neutrino magnetic moment is assumed to be $10^{-13} \, \mu_{\rm B}$, the density is $10^{10}$ g cm$^{-3}$, and the gradient of the electron fraction is $\mathrm{d} Y_e/\mathrm{d} r \simeq 10^{-7}$ cm$^{-1}$.
}
\label{fig:W_LL}
\end{figure}

A more detailed analysis of the numerical solution
of Eq.~(\ref{eq:evolution}) allows us to establish a relationship between
the magnetic field strength and parameters of
the medium in the supernova envelope, on the one
hand, and the survival probability of active neutrinos
$W_{LL}$, on the other hand. Using typical scales of
parameters in the region under consideration, see~\cite{Bethe:1990,Buras:2005},
\begin{equation}
\frac{\mathrm{d} Y_e}{\mathrm{d} r} \sim 10^{-7} \, \textrm{cm}^{-1} \,, 
\quad
\rho \sim 10^{10} \, \textrm{g cm}^{-3} \,,
\label{eq:param}
\end{equation}
we find an approximation formula,
\begin{eqnarray}
&& \frac{B_{\perp} (t)}{B_e} = f (W_{LL}) \left( 
\frac{10^{-13} \mu_{\rm B}}{\mu_\nu} \right)\times
\nonumber\\
&& \times
\left( \frac{\rho (t)}{10^{10} \, \textrm{g cm}^{-3}}
\right)^{1/2} 
\left( \frac{\mathrm{d} Y_e}{\mathrm{d} r} (t) \times 10^7 \, \textrm{cm}
\right)^{1/2} .
\label{eq:connec_B_W}
\end{eqnarray}
Here, the factor
\begin{equation}
f (W_{LL}) = 0.88 \, \frac{(1 - W_{LL})^{0.62}}{(W_{LL})^{0.13}} 
\label{eq:f_W}
\end{equation}
characterizes the degree of adiabaticity of the conversion
process. The literal adiabaticity corresponds to
the limit $f \to \infty$, when $W_{LL} \to 0$; in this case, the left-handed
neutrinos are completely converted into right-handed
ones, $W_{LR} = (1 - W_{LL}) \to 1$.

The conservative value of $10^{-13} \mu_{\rm B}$ introduced in
Eq.~(\ref{eq:connec_B_W}) as the scale for the neutrino magnetic moment
was chosen to be an order of magnitude smaller than
limit~(\ref{eq:mu_lim_summ}), so that the conversion of sterile neutrinos
produced in the supernova core through the first of the
above mechanisms into active ones did not distort the
supernova explosion dynamics. Thus, we can use the
parameters of the explosion model without allowance
for the influence of the neutrino magnetic moment.
Our analysis based on detailed data on the radial
distributions and time evolution of physical properties
in a supernova core obtained in the specific model of a
successful explosion~\cite{Janka:2009} showed
that the gradient of the electron fraction $\mathrm{d} Y_e / \mathrm{d} r$ in
Eq.~(\ref{eq:connec_B_W}) grows fairly rapidly with time at point $Y_e = 1/3$ and, thus, the envelope becomes more transparent
to active neutrinos at a fixed magnetic field strength. This means that the neutrino signal from
the supernova can be attenuated within some limited
time interval after its explosion.
Thus, if the Dirac neutrino had a magnetic moment
and if the magnetic field in the supernova envelope
were sufficiently strong, then the characteristic
effect of a significant attenuation of the initial neutrino
signal intensity peak predicted by supernova models
could take place. For example, there would be a tenfold
reduction in the neutrino signal ($W_{LL} = 0.1$) for
typical parameters of the medium at a magnetic field
strength
\begin{eqnarray}
&& B_{\perp} = 4.9 \times 10^{13} \, \textrm{Ãñ} \left( 
\frac{10^{-13} \mu_{\rm B}}{\mu_\nu} \right)\times
\nonumber\\
&& \times
\left( \frac{\rho}{10^{10} \, \textrm{g cm}^{-3}}
\right)^{1/2} 
\left( \frac{\mathrm{d} Y_e}{\mathrm{d} r} \times 10^7 \, \textrm{cm}
\right)^{1/2} .
\label{eq:B_10}
\end{eqnarray}
Note that the possible strengths of a magnetic
field generated in a supernova envelope are believed to reach $10^{16}$\, G~\cite{Bisnovatyi-Kogan:1970,Bisnovatyi-Kogan:1989,Bocquet:1995,Spruit:1999,Cardall:2001, Bisnovatyi-Kogan:2005}.

\section{The neutrino signal from SN 1987A}
\label{sec:SN1987A}

It is of interest to compare observational predictions
of the effect being discussed with the only (to
date) neutrino signal from SN 1987A, when three
underground neutrino detectors, Kamiokande-II~\cite{Hirata:1987,Hirata:1988}, 
IMB~\cite{Bionta:1987,Bratton:1988}, and the Baksan scintillation
telescope~\cite{Alexeyev:1987}, recorded electron
antineutrinos in the reaction ${\bar\nu}_e + p \to n + e^+$ for the
first time. The neutrino signal recorded by the LSD
detector~\cite{LSD:1987} 4 h 43 min earlier than
these three detectors requires a separate analysis and
we disregard it.

It should be recognized that the statistics of neutrino
events from SN 1987A is, of course, insufficient
for firm conclusions about the time evolution of the
neutrino flux to be reached: the Kamiokande, IMB,
and Baksan detectors recorded eleven (one more
event was attributed to the background), eight, and
five events (one more event was attributed to the
background), respectively. Combining the data from
all three detectors to study the time evolution of the
neutrino signal is a serious problem. First, the signal
timing accuracy was different: it was $\pm$ 50 ms for
IMB, for $\pm$ 1 min for Kamiokande, and $\pm$ 2 s for the
Baksan telescope. Second, the detectors had different
sensitivity thresholds: at an energy of positrons below
27 MeV, the efficiency of their detection by the IMB
detector dropped below 30 \%; for the Kamiokande
and Baksan detectors, the corresponding thresholds
were 7 and 9 MeV.

The simplest and, at the same time, not unfounded
solution of the signal timing problem is that the first
events in the series from different detectors are declared
to be coincident in time and an appropriate
time shift is made (see, e.g.,~\cite{Alexeyev:1988}).
The reason for this solution is that, according to
the existing supernova explosion models (see, e.g.,
~\cite{Bethe:1990,Buras:2005}), the supernova core
luminosity through electron antineutrinos reaches its
maximum within about 30 ms after the bounce and
then decreases fairly rapidly after 200 ms. Thus, the
probability of detecting the antineutrinos produced
precisely in this time interval is maximal. Making the
first events coincident in time in all three detectors
and choosing this instant as the zero time $t = 0$, we
obtain a set of 24 events~\cite{Hirata:1988,Bratton:1988,Alexeyev:1987} in the time interval
from 0 to 12.4 s, with 17 events occurring during the
first 3 s. However, the time distribution of these events
exhibits no initial peak. Thus, it may well be that the
time evolution of the only observed neutrino signal
from SN 1987A confirms the attenuation of the initial
peak described above.

\section{Neutrino pulsar}
\label{sec:pulsar}

Note another possible interesting manifestation
of the neutrino magnetic moment. If a magnetar
with a poloidal magnetic field of $10^{14} - 10^{15}$\,G is
formed during a supernova explosion, then, given that
Eqs.~(\ref{eq:evolution}) and (\ref{eq:connec_B_W}) contain the transverse magnetic field
component $B_{\perp}$, the neutrinos can avoid the conversion
of their helicity only in a narrow region near the
poles. When the nascent magnetar rotates around
an axis that does not coincide with its magnetic
moment and if we are lucky with the orientation of the
rotation axis, the neutrino signal will have a pulsating
behavior, as is illustrated in Fig. 2, i.e., a kind of a
neutrino pulsar can be observed.

\begin{figure}
\begin{center}
\includegraphics*[width=0.8\textwidth]{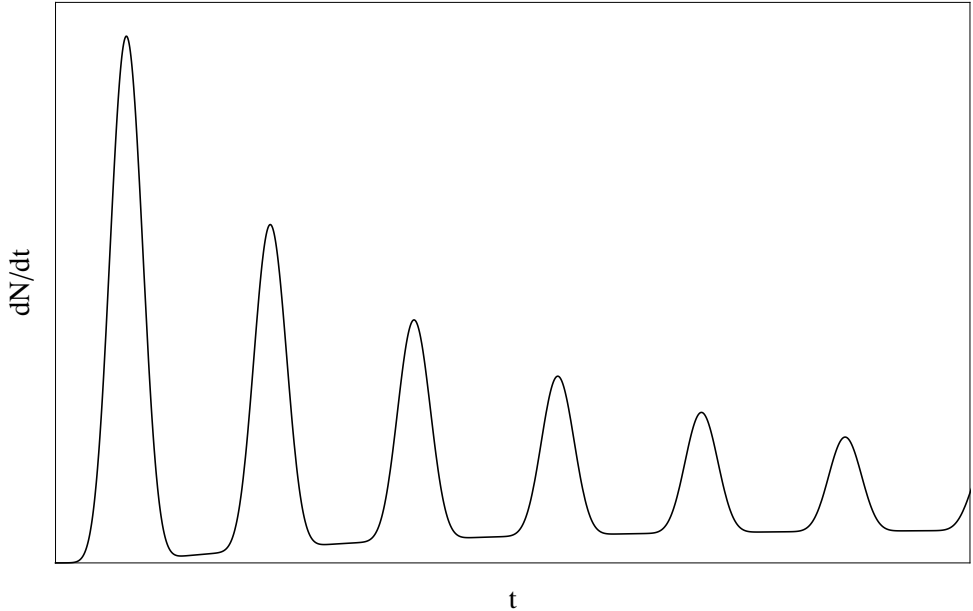}
\end{center}
\caption{Illustration of the pulsating behavior of the neutrino signal from a nascent magnetar rotating around an axis that does not coincide with its magnetic moment, a neutrino pulsar.
}
\label{fig:nu-pulsar}
\end{figure}

It should be noted that, strictly speaking, the described
influence of a strong magnetic field when the
neutrino has a magnetic moment on the time evolution
of the neutrino signal is incomplete without
allowance for the effects of neutrino flavor oscillations
(see, e.g.,~\cite{Kneller:2008}). The combined action
of these effects on the neutrino flux requires a special
study.

\section{Discussion}
\label{sec:Discussion}

We showed that if the Dirac neutrino had a magnetic moment
and if the magnetic field in the supernova
envelope were sufficiently strong, then the characteristic
effect of a significant attenuation of the initial
neutrino signal intensity peak predicted by supernova
models could take place. For instance, at typical parameters
of the medium and at a neutrino magnetic
moment of $\sim 10^{-13} \mu_{\rm B}$, i.e., an order of magnitude
smaller than the existing astrophysical limit, there
would be a tenfold reduction in the neutrino signal
even at a magnetic field strength of the order of the
critical one $B_e$.

Remarkably, as our analysis showed, the time
evolution of the only observed neutrino signal from
SN 1987A may confirm this attenuation of the initial
neutrino peak.

If a magnetar with a poloidal magnetic field is
formed in a supernova explosion, then the neutrino
signal will have a pulsating behavior, i.e., a kind of
a neutrino pulsar can be observed, when it rotates
around an axis that does not coincide with its magnetic
moment and when the orientation of its rotation
axis is ``lucky''.



\section*{Acknowledgements}

A.K. and N.M. express their deep gratitude to the organizers of the 
Seminar ``Quarks-2010'' for warm hospitality. 
We are grateful to Hans-Thomas Janka, Lorenz H\"udepohl and Bernhard M\"uller
for providing us with detailed data on radial distributions and time evolution
of physical parameters in the supernova core, obtained in their
model of supernova explosion and proto-neutron star cooling.
We thank A.A. Gvozdev and I.S. Ognev for useful remarks. 

This study was carried out within
the framework of the ``Scientific and Scientific-Educational Personnel of Innovational Russia'' Federal
Goal-Oriented Program for 2009-2013 (State
contract no. P2323) under partial financial support
by the Ministry of Education and Science of the
Russian Federation according to the ``Development
of Scientific Potential of Higher School'' Program
(project no. 2.1.1/510).


\end{document}